# DESIGNING HARDWARE/SOFTWARE SYSTEMS FOR EMBEDDED HIGH-PERFORMANCE COMPUTING

*Mário P. Véstias†, Rui Policarpo Duarte‡, Horácio C. Neto⋆*

†INESC-ID, ISEL - Instituto Superior de Engenharia de Lisboa, Instituto Politécnico de Lisboa
‡ Departamento de Ciência e Tecnologia, Universidade Autónoma de Lisboa
⋆INESC-ID, Instituto Superior Técnico, Universidade de Lisboa, Portugal,
mvestias@deetc.isel.ipl.pt, rpduarte@ual.pt, hcn@inesc-id.pt

## ABSTRACT

In this work, we propose an architecture and methodology to design hardware/software systems for high-performance embedded computing on FPGA. The hardware side is based on a many-core architecture whose design is generated automatically given a set of architectural parameters. Both the architecture and the methodology were evaluated running dense matrix multiplication and sparse matrix-vector multiplication on a ZYNQ-7020 FPGA platform. The results show that using a system-level design of the system avoids complex hardware design and still provides good performance results.

## I. INTRODUCTION

Computing requirements of embedded systems are rapidly increasing with stringent real-time requirements, together with low power and low cost. Single processor solutions are unable to provide the required performance and at the same time keep the power consumption low. Hardware/software architectures where the most computational demanding parts of the application run in dedicated hardware have shown very good performance, area and power efficiencies.

While efficient, hardware/software architectures are in general difficult to obtain since designing dedicated hardware for a specific algorithm in FPGAs requires hardware expertise. From the perspective of the software programmer, an automatic flow to design and configure the hardware/software architecture is essential.

In this work, our approach is to use a configurable hardware coprocessor whose design is generated automatically after being parameterized by the programmer. The coprocessor consists of a many-core architecture that is automatically generated and integrated with the embedded processor. The many-core coprocessor is configurable in the number of cores, the system memory (number and size of local memories, cache and interfaces to external memory) and the topology of the interconnection network (Network-on-Chip, ring or simply point-to-point connections). The number and type of arithmetic operations of each core, number formats, including floating-point and integer can also be configured. Each core has local memory, an arithmetic unit and input/output interfaces. Keeping the core simple permits to explore more parallelism, reduces power consumption and makes configuration easier. The design and programming of the architecture was integrated in a proposed design flow that starts with the algorithm specification and outputs the hardware/software system to be implemented in a SoC FPGA.

Constraining the hardware design space to a hardware template may reduce the performance compared to a fully-optimized solution. However, it typically provides a good tradeoff between hardware performance, hardware portability and design time.

The paper is organized as follows. Section 2 describes the state-of-the-art in tools and architectures to design hardware/software processing architectures for FPGAs. Section 3 describes the proposed hardware/software architecture. Section 4 describes the proposed architecture design flow. Section 5 shows the results obtained and section 6 concludes the paper.

## II. RELATED WORK

Many commercial and academic tools have been proposed to raise the design synthesis level of FPGAs and therefore reduce the design time and design efforts. High-level synthesis tools exist to generate hardware from C/C++ (C-to-Verilog [5], Catapult-C [3], Mitrion-C [4], ImpulseC [2], HandelC [1], Xilinx AutoESL, etc), SystemC (Bluespec [6], Xilinx AutoESL), Java (JHDL [7], MaxCompiler [8]), Python (MyHDL [9]), among others.

The most common approach for hardware compilation is to start with C/C++, with some language restrictions to avoid recursions and pointers. Compiler techniques were proposed to generate an optimized design for specific hardware platforms. For some large designs the generated hardware obtained with these tools were able to achieve better optimized hardware implementations compared to hand-made solutions.

In these tools, additional annotations are used in the code to control some implementation options. The compil-ers extract as much as possible instruction-level parallelism that can be exposed using techniques like loop unrolling and pipelining. These compilers automatically generate the hardware but programmers must be aware of the hardware programming model which requires some knowledge on circuit design.

Another research direction for hardware design consists of using overlays that implement an intermediate reconfigurable architecture within the user logic of the FPGA. In [10] and [12] programmable overlays are used to increase the performance of DSP workloads on FPGA. In [11] the viability of a GPU-like overlay for FPGA was analyzed. However, whether GPU-like programming models and architectures are a good way to design many-cores on FPGA is yet to be checked. Also, if not carefully designed these overlays will run with low sustained performances compared to their peak performances.

Our proposal for the design of hardware/software high-performance embedded systems is to consider the hardware side as a many-core coprocessor. The coprocessor architecture is configurable at a system-level where the programmer only has to specify system-level parameters, like, the number of cores, the numerical precision of the arithmetic units, among others. Each core runs from simple to complex arithmetic operations, like vector multiplication, matrix multiplication. These operations are part of a library and new operations can be added through microcode programming.

A few many-core designs on FPGA have already been proposed. The MPLEM system [13] consists of Xilinx MicroBlaze soft-core processors connected with On-chip Peripheral Bus (OPB) buses. In [14] a system with 24 MicroBlaze cores interconnected with an Arteris NoC [15] was proposed. The system was implemented in a Virtex-4 FX-140 FPGA.

Tumeo et al. [16] proposed a real-time many-core system for automotive applications also based on Microblaze. Each core contains local data memory and all cores share an external RAM for shared data and instructions. Cores can communicate through a bus-based shared memory, or a message-passing subsystem built upon a crossbar module.

HeMPS-based systems [17] are homogeneous multi-processor platforms using a network-on-chip (NoC) interconnection. Each processing element has a Plasma processor [19], an internal RAM block, a network interface to the NoC and a DMA engine. The platform is automatically generated and the number of processors can be customized. Design space exploration is based on simulation. Processors are modeled using cycle accurate instruction set simulators and local memories with C/SystemC models.

MARC (Many-core Approach to Reconfigurable Computing) [18] is a many-core template comprising one control processor and multiple processors for running tasks as SIMD (single instruction multiple data) units. Cores can be configured as RISC processors or synthesized as full-custom datapaths. Each core has local private memory and have access to an internal shared memory. Processors are interconnected with a network selected from a library with various topologies, including crossbar and torus.

SMYLEref [21] is a many-core architecture for embedded systems prototyped in FPGA. The architecture consists of multiple clusters arranged in a two-dimensional array connected with a NoC. Each cluster has a number of scalar processors connected with a local bus. Each core has dedicated instruction and data L1 caches. A second layer of cache exists in each cluster shared by all cores. The processor core is a Geyser [22].

Most of these many-core proposals rely on general-purpose embedded processors as the core unit. This increases flexibility but decreases performance and area efficiency. In approaches, like MARC, it is possible to customize the processor with a dedicated datapath that requires hardware design, but the results are still far from the peak capacity of the FPGA. Design space exploration is not specified in most approaches, but HeMPS, for example, uses ISS and system level simulation models to explore different platforms.

In our architecture, the core elements are based on simple processing units with reduced control, small local memories and arithmetic units. Each core unit can be individually configured in terms of local memory size and number and type of arithmetic operations. This permits to improve performance and area efficiency when compared to many-core architectures based on general-purpose embedded processors. We consider a customizable interconnection network that can be a bus, a crossbar, a NoC or a ring, and that can use point-to-point connections and/or a mix of these topologies. We rely on SystemC to do the design space exploration. We model the many-core platform and the algorithm using SystemC and do system level simulations to help in design space exploration.

### III. HARDWARE/SOFTWARE MANY-CORE ARCHITECTURE

The proposed hardware/software architecture consists of an embedded processor and the many-core architecture (see figure 1).

The many-core has access to external memory through a DMA that is configured by the embedded processor. The DMA is responsible for sending/receiving data to/from memory and for forwarding this data to the network. In order to improve the bandwidth when requesting elements stored non-sequentially in memory, the DMA has a cache to buffer bursts of data and thus enable faster access. Each time non-sequential data is requested from memory, a burst of sequentially-stored elements is fetched (cacheline size). The first element of the burst is the data requested. This data is immediately forwarded to the processors. The other elements are stored in cache.

The cores are organized in clusters. Each cluster has a local lite processor (local PE) to program the cores and

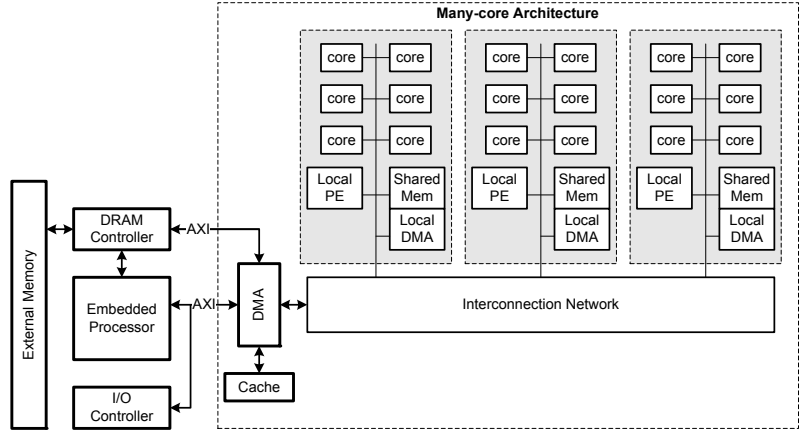

**Fig. 1**. Hardware/software many-core architecture

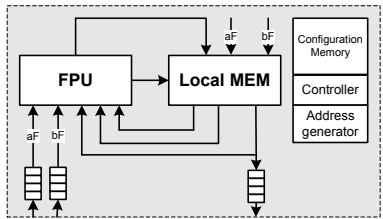

**Fig. 2**. Architecture of the core unit

the local DMA and control data communication; shared memory and a local DMA to transfer data to/from the cores. Cores are connected to the communication network through input and output buffers. Input buffers are connected to the FPU and to the local memory (*aF* and *bF* signals)

Each core has an arithmetic unit and a local data memory (see Figure 2). The arithmetic unit can be statically or dynamically configured to execute a set of basic functions: add/sub, multiplier, fused multiply-add, reciprocal, square root and inverse square-root [20]; and a set of more complex functions: vector multiplication, block matrix multiplication, etc. Each core can be configured with a different combination of operations and new complex operations can be added trough microprogramming. The local memory is implemented with dual-port block RAMs that are used to store temporary variables (registers are implemented with this memory), coefficients to implement some of the arithmetic operators, constants, and output data.

## IV. ARCHITECTURE DESIGN FLOW

A many-core generator was developed to automatically generate the many-core architecture from a set of architecture specifications. Also, an instance of the many-core platform is also automatically modeled in SystemC for a system level simulation to determine the number of execution cycles considering different configurations of the architecture. In this version, the code to run on the cores and to program the DMA are obtained manually (see figure 3).

The flow starts with the configuration of the architec-ture. The architecture may be simulated at system-level for a specific algorithm. To do this, a SystemC model of the architecture is automatically generated and then the algorithm must be parallelized, modelled in SystemC together with the architecture description. After these steps, the many-core architecture is generated using a library of cores and the software to run in the DMA and in the cores is manually generated by the designer. Cores are programmed with microcode instructions. The many-core is then exported to XPS (Xilinx Platform Studio) and integrated with the embedded processor present in the ZYNQ platform.

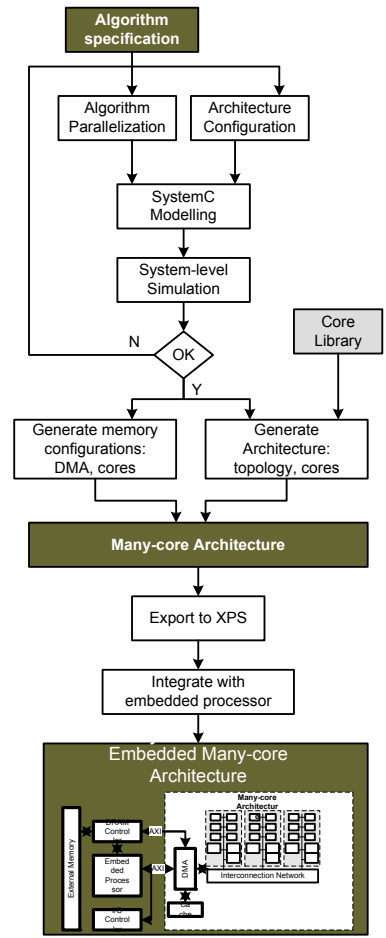

**Fig. 3**. Many-core design flow

The design space exploration process is manual, that is, the designer is responsible for manually specifying different configurations of the many-core architecture and different algorithm parallelizations. The only automatic processes are the generation of a SystemC description of the architecture given a particular configuration of the architecture and algorithm, and the generation of a VHDL description of the many-core architecture to be synthesized and its integration in the XPS from Xilinx to generate the complete hardware/software embedded architecture. Along the flow the designer does not have to design hardware since the hardware is generated automatically.

## V. RESULTS

To evaluate the flow and the architecture, we have considered parallel algorithms for dense matrix multiplication and sparse matrix-vector multiplication. For both, we explored the design space looking for the best many-core using the proposed flow. Both architectures were implemented in a ZYNQ-7000 SoC XCZ7020-CLG484 and tested on a ZedBoard with this device.

### V-A. Configuration of the Architecture for Dense Matrix Multiplication

Matrix multiplication $C = A \times B$ is implemented as a parallel block matrix algorithm that partitions $C$ matrix into smaller sub-matrices (blocks) and works with these blocks. All matrices are square and have the same size ($n \times n$).

The $C$ matrix is divided in blocks with size $n \times xp$. Each of these blocks is calculated by p cores simultaneously. Each core is responsible for a sub block with size $n \times x$ which in turn is divided in smaller blocks with dimension $y \times x$. The size of these smaller blocks, $C_{ij}$, depends on the local memory size. To generate a block $C_{ij}$ the processor multiplies a block $y \times n$ from matrix A with a block $n \times x$ from matrix B. The multiplication is implemented as a sequence of k partial block multiplications,

$$C_{ij} = \sum_{k=1}^{k_0} A_{ik} \times B_{kj} \quad (1)$$

Each partial block is the multiplication of a $y \times z$ sub block $A_{ik}$ with a $z \times x$ sub block $B_{kj}$, resulting in a partial sub block $C_{ij_k}$ of size $y \times x$. The final $C_{ij}$ result is obtained after accumulating the $k$ partial block results.

The partial block multiplications are implemented as follows. First, each core receives and stores its $B_{qj}$ elements. Then, $A_{iq}$ elements are broadcasted to all cores. As the $A_{iq}$ elements arrive, they are multiplied by all $B_{qj}$ elements stored in local memory. The partial results of each block $C_{ij}$ are also stored in local memory. In the final iteration, the elements of the result block $C_{ij}$ are sent to the external memory. As referred, the local memory in each processor must store the blocks of $B$ (size $z \times x$) and $C$ (size $x \times y$) under processing.

At the algorithmic level, $x$, $y$ and $z$ are variables and thus different performance results are obtained by changing these values. To optimize the final solution, we have considered the theoretical results in [25] to determine these values. According to the referenced theoretical results, the number of communications with the external memory does not depend on the dimension $z$ of the sub blocks. Therefore, $z$ can be simply made equal to 1 in order to reduce the local memory required. The local memory necessary to store the sub blocks of $B$ (size $1 \times x$) is doubled in order to enable the processor to store a new $B$ sub-block while still performing the computations with the former $B$ sub-block.

Also according to this reference, the dimensions of the sub blocks $C_{ij}$ that minimize the number of communications, as a function of the available local memory $L$, are

$$x = \frac{L}{2 + \sqrt{pL}} \quad y = \sqrt{pL} \quad (2)$$

At the architectural level, matrix multiplication requires multiply and add operations. So, the arithmetic units of all cores are configured as fused multiply-add. We have configured the many-core with 16 and 32 cores, all with the same local memory size and a DMA cache with support for up to 16 cachelines.

Assuming an architecture with 32 KBytes of local memory for the 16-core and 16 KBytes for the 32-core architecture, we have determined the utilization of resources and the number of execution cycles (see table I). Both architectures achieve high performance efficiencies (peak performance/measured performance), 86% and 84%, respectively. The 16-core achieves 7 GFLOPs and the 32-core achieves 13.4 GFLOPs.

**Table I**. Results for matrix multiplication

|              | Core  | Arch. 16-cores | Arch. 32-cores |
|--------------|-------|----------------|----------------|
| LUTs         | 1,364 | 24,390         | 46,576         |
| DSPs         | 4     | 71             | 135            |
| BRAMs        | 8/4   | 140            | 140            |
| Freq. (MHz)  | 250   | 250            | 250            |
| Cycles       | —     | 77,772,668     | 39,796,887     |
| Time (s)     | —     | 0.31           | 0.16           |
| GFLOPs       | —     | 7              | 13.5           |
| Peak GFLOPs  | 0.5   | 8              | 16             |
| Efficiency   | —     | 86%            | 84%            |

Compared to previous implementations, ours has about half of the performance of the dedicated architecture for matrix multiplication in [24], but consumes only about 25% of the resources. Doubling the number of cores of our architecture would provide an architecture with the same performance, assuming enough memory bandwidth. In terms of efficiency, our architecture is better. We also have higher efficiencies compared to the dedicated many-core proposed in [23].

### V-B. Configuration of the Architecture for Sparse Matrix Multiplication

We have parallelized the sparse matrix-vector multiplication algorithm to run in a many-core architecture. In this paper, we briefly describe the parallelization process (See [28] for a detailed explanation).

Sparse matrix-vector multiplication is the mathematical operation given by

$$y = A \times x$$

where matrix $A$ is a sparse matrix, $x$ is the input vector and $y$ the result of the product between $A$ and $x$. Given a matrix $A$ of size $n \times m$, vector $x$ is necessarily of size $1 \times m$ and vector $y$ of size $n \times 1$.

A matrix is typically stored as a two-dimensional array. Each entry in the array represents an element $A_{i;j}$ of the matrix and is accessed by the two indices *i* and *j*. Conventionally, *i* is the row index, numbered from top to bottom, and *j* is the column index, numbered from left to right. For an $M \times N$ matrix, the amount of memory required to store the matrix in this format is proportional to $M \times N$ (disregarding the fact that the dimensions of the matrix also need to be stored).

In the case of a sparse matrix, substantial memory requirement reductions can be realized by storing only the non-zero entries. Depending on the number and distribution of the non-zero entries, different data structures can be used and yield huge savings in memory when compared to the basic approach. In this work we have used Compressed Sparse Column (CSC). The compressed sparse column format stores an initial sparse $M \times N$ matrix A in column form using three one-dimensional arrays.

Work attribution to cores was made by nonzero indexes. This means that instead of row ranges, single rows were attributed to each processor. This attribution is done in a round-robin fashion.

To show that the work scheduling to processors in the previous row assignment is balanced, tests using a data set of matrices were run and the percentage of nonzeros assigned to each processor was measured. Results indicate that, for a system composed of four processors, the load balancing measured by percentage of the total number of nonzeros is around 25% for each processor, guaranteeing a good work load balance.

In our design all necessary data to perform a sparse matrix-vector multiplication is in external memory. Therefore, the system implemented reads all data from external memory and writes the result back to external memory. The algorithm is scalable to any number of cores.

The *DMA* module is responsible for moving data between the external memory and the cores. The DMA is controlled by micro instructions provided by the ARM processor located in the Processing System through an AXI General Purpose interface. The DMA unit is structured in two independent modules which enable it to process read and write operation simultaneously. Each core is composed of an input buffer, a Fused Multiplier-Adder (FMA) and local memory.

Table II represents the performance results obtained for the sparse matrix-vector hardware implementation working at an operating frequency of 100 MHz with two cores. With the available bandwidth using more processors improves marginally the execution time.

**Table II**. Performance results of the proposed architecture

| Test name | Maragal_2 | flower_5_4 | BIBD_14_7 | LD_pilot87 |
|---|---|---|---|---|
| NNZ | 4357 | 43942 | 72072 | 74949 |
| M | 555 | 5226 | 91 | 2030 |
| NNZ per Col | [0, 139] | [1, 3] | [21, 21] | [1, 96] |
| ARM exec (us) | 128 | 1644 | 2055 | 2222 |
| HW exec (us) | 94 | 1077 | 1438 | 1647 |
| HW/ARM | 1,18 | 1,31 | 1,43 | 1,35 |

Each column corresponds to a different test with different matrix and vector inputs. *NNZ* stands for the number of non-zero elements and *M* is then number of rows in the input matrix. We also specify the number of non-zeros per column (*NNZ per Col*).

We have extrapolated our system to determine its performance for different memory bandwidths and compared to previous works ([26], [29], [30], [31], [32] and [27]). The average efficiencies determined are across different input matrices. Our work, presents superior efficiencies (from 44% to 66% on average) in all cases except when comparing to [27] (90% efficiency) and [26] (80% of efficiency). However, the efficiencies presented in [26] are based on a different algorithmic solution and for very specific matrices and the efficiencies presented in [27] are theoretical without taking into consideration limitations from architectural structures, like memory bandwidth, that cannot be obtained as ideally assumed in the model.

### VI. CONCLUSION

A configurable hardware/software architecture for high-performance embedded computing was proposed. The many-core architecture is configurable at system-level. A design flow to automatically generate the hardware/software architecture was also proposed that starts with the configuration of the architecture and ends with the implementation targeting an FPGA.

Previous proposals of many-core architectures for embedded systems are based on general-purpose embedded processors. Compared to our many-core, these systems in general have a better support to run control intensive kernels or threads but are less efficient for data intensive applications in terms of performance and area. This is because our cores are simpler and application optimized, and can also support higher operating frequencies.

We have evaluated the architecture for parallel dense matrix multiplication and sparse matrix-vector multiplication. The results show that the architectures generated achieves performances close to those of state-of-the-art dedicated circuits and performance efficiencies near 90% without requiring hardware expertise to design the many-core architecture.